\documentclass{llncs}
\usepackage{epsf}
\usepackage{etex}
\usepackage{amssymb}
\usepackage{amsmath,epsfig}
\usepackage{color}
\usepackage{multirow}
\usepackage{float}
\usepackage{pgfplots}
\usepackage{subfigure}
\usepackage{amsmath}
\usepackage{pstricks-add}
\usepackage{pstricks}
\usepackage{url}

\def\x{{\mathbf x}}

\begin{document}

\title{Automatic Text Summarization Approaches\\to Speed up Topic Model Learning Process}

\author{Mohamed Morchid$^1$ \and Juan-Manuel Torres-Moreno$^{1,2}$ \and  Richard Dufour$^1$ \and Javier Ram\'irez-Rodr\'iguez$^4$ \and Georges Linar\`es$^1$}

\institute{
$^1$LIA - Universit\'e d'Avignon et des Pays de Vaucluse (France)\\
\{firstname.lastname\}@univ-avignon.fr\\
$^2$\'Ecole Polytechnique de Montr\'eal, Qu\'ebec (Canada) \\
$^3$Universidad Aut\'onoma Metropolitana Azcapotzalco (Mexico)\\jararo@azc.uam.mx}

\titlerunning{Automatic Text Summarization Approaches}
\authorrunning{Morchid et al.}

\maketitle

\begin{abstract}
The number of documents available into Internet moves each day up. 
For this reason, processing this amount of information effectively and expressibly becomes a major concern for companies and scientists. 
Methods that represent a textual document by a topic representation are widely used in Information Retrieval (IR) to process big data such as Wikipedia articles. 
One of the main difficulty in using topic model on huge data collection is related to the material resources (CPU time and memory) required for model estimate. 
To deal with this issue, we propose to build topic spaces from summarized documents. 
In this paper, we present a study of topic space representation in the context of big data. The topic space representation behavior is analyzed on different languages. 
Experiments show that topic spaces estimated from text summaries are as relevant as those estimated from the complete documents. The real advantage of such an approach is the processing time gain: we showed that the processing time can be drastically reduced using summarized documents (more than 60\% in general). 
This study finally points out the differences between thematic representations of documents depending on the targeted languages such as English or 
latin languages.\footnote{Preprint of \textsl{International Journal of Computational Linguistics and Applications}, 7(2):87-109, 2016.}
\end{abstract}

\section{Introduction}
\label{sec:introduction}

The number of documents available into Internet moves each day up in an exponential way.  For this reason, processing this amount of information effectively and expressibly becomes a major concern for companies and scientists. An important part of the information is conveyed through textual documents such as blogs or micro-blogs, general or advertise websites, and encyclopedic documents. This last type of textual data increases each day with new articles, which convey large and heterogenous information. The most famous and used collaborative Internet encyclopedia is Wikipedia, enriched by worldwide volunteers. It is the 12$^{th}$ most visited website in the USA, with around 10.65 million users visiting the site daily, and a total reaching 39 millions of the estimated 173.3 million Internet users in the USA\footnote{http://www.alexa.com}~\footnote{http://www.metrics2.com}. 

The massive number of documents provided by Wikipedia is mainly exploited by Natural Language Processing (NLP) scientists in various tasks such as keyword extraction, document clustering, automatic text summarization\ldots Different classical representations of a document, such as term-frequency based representation~\cite{salton1989automatic}, have been proposed to extract word-level information from this large amount of data in a limited time. Nonetheless, these straightforward representations obtain poor results in many NLP tasks with respect to more abstract and complex representations. Indeed, the classical term-frequency representation reveals little in way of intra- or inter-document statistical structure, and does not allow us to capture possible and unpredictable context dependencies. For these reasons, more abstract representations based on latent topics have been proposed. The most known and used one is the latent Dirichlet allocation (LDA)~\cite{blei2003latent} approach which outperforms classical methods in many NLP tasks. The main drawback of this topic-based representation is the time needed to learn LDA latent variables. This massive waste of time that occurs during the LDA learning process, is mainly due to the documents size along with the number of documents, which is highly visible in the context of big data such as Wikipedia. 

The solution proposed in this article is to summarize documents contained into a big data corpus (here Wikipedia) and then, learn a LDA topic space. This should answer the these three raised difficulties:
\begin{enumerate}
\item[$\bullet$] reducing the processing time during the LDA learning process,
\item[$\bullet$] retaining the intelligibility of documents,
\item[$\bullet$] maintaining the quality of LDA models.
\end{enumerate} 

With this summarization approach, the size of documents will be drastically reduced, the intelligibility of documents will be preserved, and we make the assumption that the LDA model quality will be conserved. Moreover,  for all these reasons, the classical term-frequency document reduction is not considered in this paper. Indeed, this extraction of a subset of words to represent the document content allows us to reduce the document size, but does not keep the document structure and then, the intelligibility of each document.



The main objective of the paper is to compare topic space representations using complete documents and summarized ones. The idea behind is to show the effectiveness of this document representation, in terms of performance and time-processing reduction,  when summarized documents are used. The topic space representation behavior is analyzed on different languages (English, French and Spanish).
In the series of proposed experiments, the topic models built from complete and summarized documents are evaluated using the Jensen-Shannon ($\mathcal{JS}$) divergence measure as well as the perplexity measure. 
To the best of our knowledge, this is the most extensive set of experiments interpreting the evaluation of topic spaces built from complete and summarized documents without human models.

The rest of the paper is organized in the following way: first, Section~\ref{sec:relatedWork} introduces related work in the areas of topic modeling and automatic text summarization evaluations.
Then, Section~\ref{s:proposedApproach} describes the proposed approach, including the topic representation adopted in our work and the different summarization systems employed. 
Section~\ref{sec:SQM} presents the topic space quality measures used for the evaluation. Experiments carried out along with with the results  presented in Section~\ref{sec:Experiments}.
A discussion is finally proposed in Section~\ref{sec:Discussion} before concluding in~Section~\ref{sec:conclusionsAndFutureWork}.   

\section{Related work}
\label{sec:relatedWork}

Several methods were proposed by Information Retrieval (IR) researchers to process large corpus of documents such as Wikipedia encyclopedia. 
All these methods consider documents as a bag-of-words~\cite{salton1989automatic} where the word order is not taken into account. 

Among the first methods proposed in IR,~\cite{baeza1999modern} propose to reduce each document from a discrete space (words and documents) to a vector of numeral values represented by the word counts (number of occurrences) in the document named TF-IDF~\cite{salton1983introduction}. This approach showed its effectiveness in different tasks, and more precisely in the basic identification of discriminative words for a document~\cite{salton1973specification}. However, this method has many weaknesses such as the small amount of reduction in description length, or the weak of inter- or intra-statistical structure of documents in the text corpus. 

To substantiate the claims regarding TF-IDF method, IR researchers have proposed several other dimensionality reductions such as Latent Semantic Analysis (LSA)~\cite{deerwester1990indexing,bellegarda1997latent} which uses a {\it singular value decomposition} (SVD) to reduce the space dimension. 

This method was improved by~\cite{hofmann1999probabilistic} which proposed a Probabilistic LSA (PLSA). PLSA models each word in a document as a sample from a mixture model, where the mixture components are multinomial random variables that can be viewed as representations of topics. This method demonstrated its performance on various tasks, such as sentence~\cite{bellegarda2000exploiting} or  keyword~\cite{suzuki1998keyword} extraction. In spite of the effectiveness of the PLSA approach, this method has two main drawbacks. The distribution of topics in PLSA is indexed by training documents. Thus, the number of its parameters grows with the training document set size and then, the model is prone to overfitting which is a main issue in an IR task such as documents clustering. However, to address this shortcoming, a tempering heuristic is used to smooth the parameter of PLSA models for acceptable predictive performance: the authors in~\cite{popescul2001probabilistic} showed that overfitting can occur even if tempering process is used. 

To overcome these two issues, the latent Dirichlet allocation (LDA)~\cite{blei2003latent} method was proposed. Thus, the number of LDA parameters does not grow with the size of the training corpus and LDA is not candidate for overfitting. Next section describes more precisely the LDA approach that will be used in our experimental study.

The authors in~\cite{louis-nenkova:2009:EMNLP} evaluated the effectiveness of the Jensen-Shannon ($\mathcal{JS}$) theoretic measure~\cite{Lin'91a} in predicting systems ranks in two summarization tasks: {\it query-focused} and {\it update summarization}. 
 They have shown that ranks produced by {\sc Pyramids} and those produced by $\mathcal{JS}$ measure correlate. 
However, they did not investigate the effect of the measure in summarization tasks such as generic  multi-document summarization (DUC 2004 Task 2), biographical summarization (DUC 2004 Task 5), opinion summarization (TAC 2008 OS),  and summarization in languages other than English.

Next section describes the proposed approach followed in this article, including the topic space representation with the LDA approach and its evaluation with the perplexity and the Jensen-Shannon metrics.

\section{Overview of the proposed approach}
\label{s:proposedApproach}

\begin{figure}[htb]
\begin{center} 
\hspace{-15mm}
\includegraphics[scale=0.5]{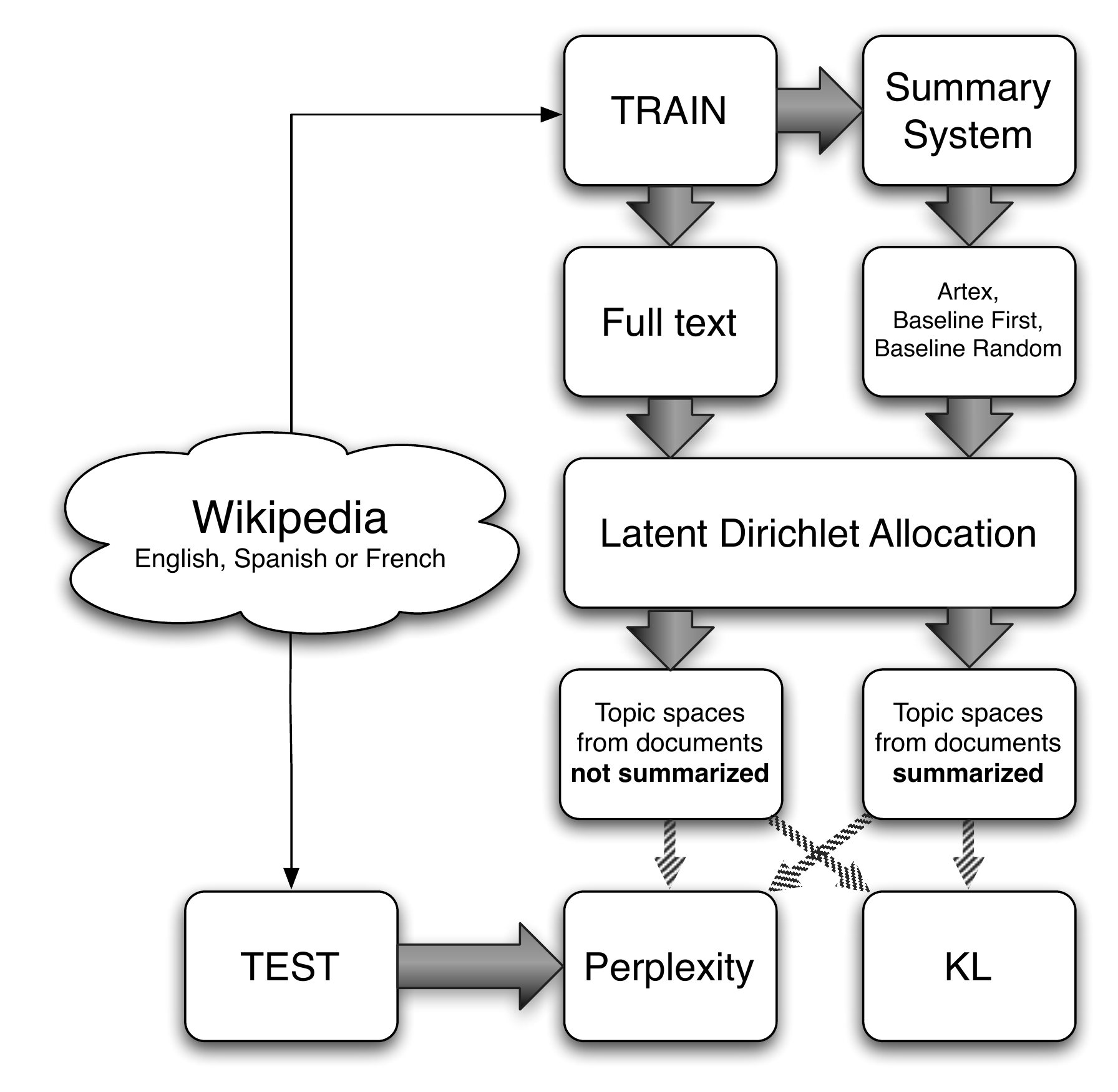} 
\caption{Overview of the proposed approach.} \label{fig:proposedApproach}
\end{center} 
\end{figure}

Figure~\ref{fig:proposedApproach} describes the approach proposed in this paper to evaluate the quality of a topic model representation with and without automatic text summarization systems.
The latent Dirichlet allocation (LDA) approach, described in details in the next section, is used for topic representation, in conjunction with different state-of-the-art summarization systems presented in Section~\ref{sec:summary}.

\subsection{Topic representation: latent Dirichlet allocation}
\label{sec:topicSpace}


LDA is a generative model which considers a document, seen as a {\it bag-of-words}~\cite{salton1989automatic}, as a mixture of latent topics. In opposition to a multinomial mixture model, LDA considers that a theme is associated to each occurrence of a word composing the document, rather than associate a topic with the complete document. Thereby, a document can change of topics from a word to another. However, the word occurrences are connected by a latent variable which controls the global respect of the distribution of the topics in the document. These latent topics are characterized by a distribution of word probabilities associated with them. PLSA and LDA models have been shown to generally outperform LSA on IR tasks~\cite{hofmann2001}. Moreover, LDA provides a direct estimate of the relevance of a topic knowing a word set.

\begin{figure}[htb]
\begin{center} 
\includegraphics[scale=0.5]{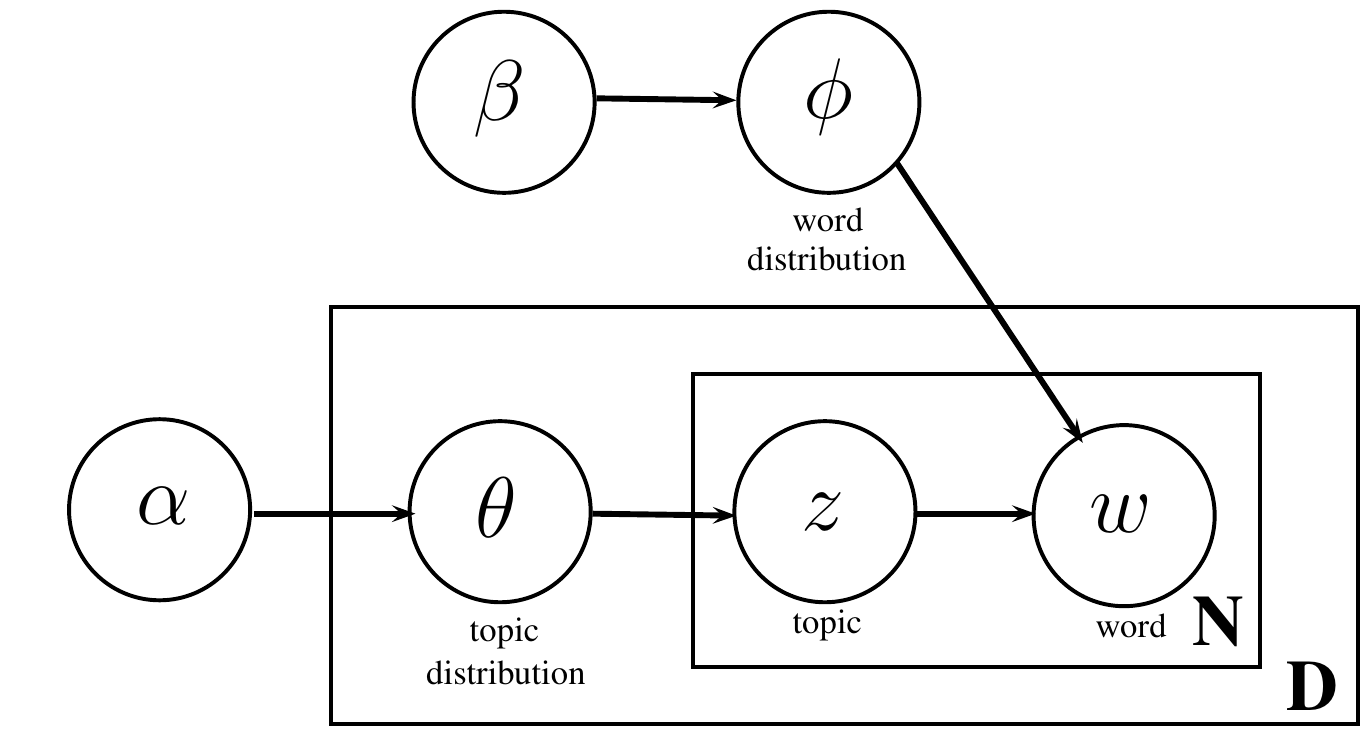} 
\caption{LDA Formalism.} \label{lda}
\end{center} 
\end{figure}

Figure \ref{lda} shows the LDA formalism. For every document $d$ of a corpus $D$, a first parameter~$\theta$ is drawn according to a Dirichlet law of parameter $\alpha$. A second parameter $\phi$ is drawn according to the same Dirichlet law of parameter $\beta$. Then, to generate every word $w$ of the document $c$, a latent topic $z$ is drawn from a multinomial distribution on $\theta$. Knowing this topic $z$, the distribution of the words is a multinomial of parameters $\phi$. The parameter $\theta$ is drawn for all the documents from the same {\it prior} parameter $\alpha$. This allows to obtain a parameter binding all the documents together~\cite {blei2003latent}.

Several techniques have been proposed to estimate LDA parameters, such as Variational Methods~\cite{blei2003latent}, Expectation-propagation~\cite{minka2002expectation} or Gibbs Sampling~\cite{griffiths2004finding}. Gibbs Sampling is a special case of Markov-chain Monte Carlo (MCMC)~\cite{geman1984stochastic} and gives a simple algorithm to approximate inference in high-dimensional models such as LDA~\cite{heinrich2005parameter}. This overcomes the difficulty to directly and exactly estimate parameters that maximize the likelihood of the whole data collection defined as: $p(W|\overrightarrow{\alpha},\overrightarrow{\beta})=\prod_{m=1}^M p(\overrightarrow{w}_m|\overrightarrow{\alpha},\overrightarrow{\beta})$ for the whole data collection $W=\{\overrightarrow{w}_m\}_{m=1}^M$ knowing the Dirichlet parameters $\overrightarrow{\alpha}$ and $\overrightarrow{\beta}$.

The first use of Gibbs Sampling for estimating LDA is reported in~\cite{griffiths2004finding} and a more comprehensive description of this method can be found in~\cite{heinrich2005parameter}. 




The next section  describes the income of the LDA technique. The input of the LDA method is an automatic summary of each document of the train corpus. These summaries are built with different systems.

\subsection{Automatic Text Summarization systems}
\label{sec:summary}
Various text summarization systems have been proposed over the years \cite{torres-moreno2014}. 
Two baseline systems as well as the ARTEX summarization system, that reaches state-of-the-art performance~\cite{artex}, are presented in this section. 

\subsubsection{Baseline first (BF)}
\label{subsec:BF}

The {\it Baseline first} (or leadbase) selects the $n$ first sentences of the documents, where $n$ is determined by a \textsl{compression rate}. Although very simple, this method is a strong baseline for the performance of any automatic summarization system~\cite{ledeneva2008terms,manning:99}. This very old and very simple sentence weighting heuristic does not involve any  terms at all: it assigns highest weight to the first sentences of the text. Texts of some genres, such as news reports or scientific papers, are specifically designed for this heuristic: {\it e.g.}, any scientific paper contains a ready summary at the beginning. This gives a baseline~\cite{duc2002} that proves to be very hard to beat on such texts. It is worth noting that in Document Understanding Conference (DUC) competitions~\cite{duc2002} only five systems performed above this baseline, which does not demerit the other systems because this baseline is genre-specific. 

\subsubsection{Baseline random (BR)}
\label{subsec:BR}

The {\it Baseline random}~\cite{ledeneva2008terms} randomly selects $n$ sentences of the documents, where $n$ is also determined by a \textsl{compression rate}.
This method is the classic baseline for measuring the performance of automatic text summarization systems.   

\subsubsection{ARTEX}
\label{subsec:ARTEX}

\begin{figure}[!h]
\begin{center}
\includegraphics[scale=0.4]{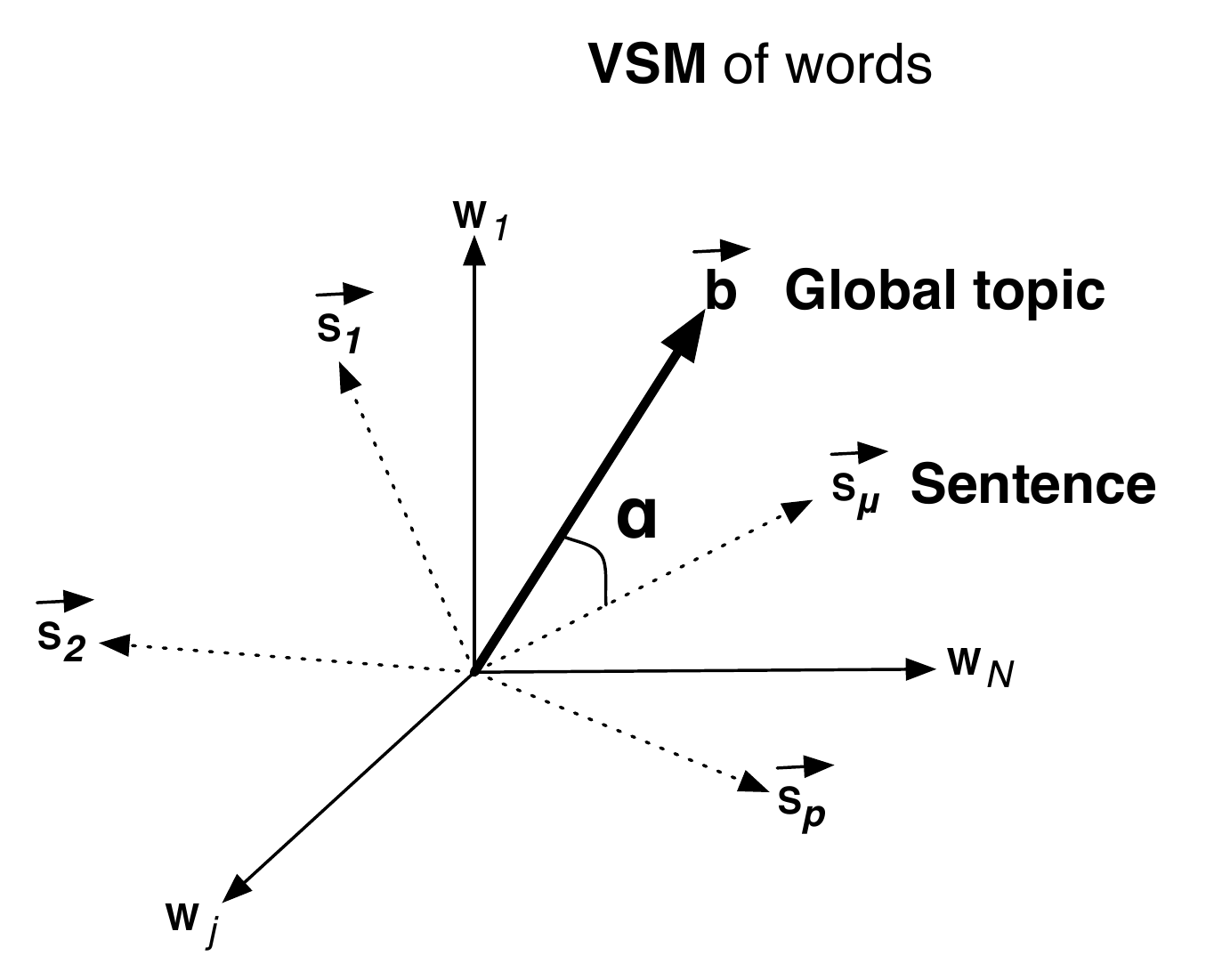}
 \caption{The ``global topic'' in a Vector Space Model of $N$ words.}
 \label{fig:artex_sentence}
\end{center}
\end{figure}

{\it {\bf A}nothe{\bf R} {\bf TEX}t} (ARTEX) algorithm~\cite{artex} is another simple extractive algorithm. The main idea is to represent the text in a suitable space model (VSM). Then, an average document vector that represents the average (the ``global topic'') of all sentence vectors is constructed. At the same time, the ``lexical weight'' for each sentence, {\it i.e.} the number of words in the sentence, is obtained. After that, the angle between the average document and each sentence is calculated. Narrow angles $\alpha$ indicate that the sentences near the ``global topic'' should be important and are therefore extracted. See Figure~\ref{fig:artex_sentence} for the VSM of words: $p$ vector sentences and the average ``global topic'' are represented in a $N$ dimensional space of words. The angle $\alpha$ between the sentence $\overrightarrow{s_{\mu}}$ and the global topic $\overrightarrow{b}$ is processed as follow:
\begin{equation}
cos(\alpha)=\frac{\overrightarrow{b}\times \overrightarrow{s_\mu}}{||\overrightarrow{b}||.||\overrightarrow{s_\mu}||}
\end{equation}

\begin{figure}[!h]
\begin{center}
\includegraphics[scale=0.4]{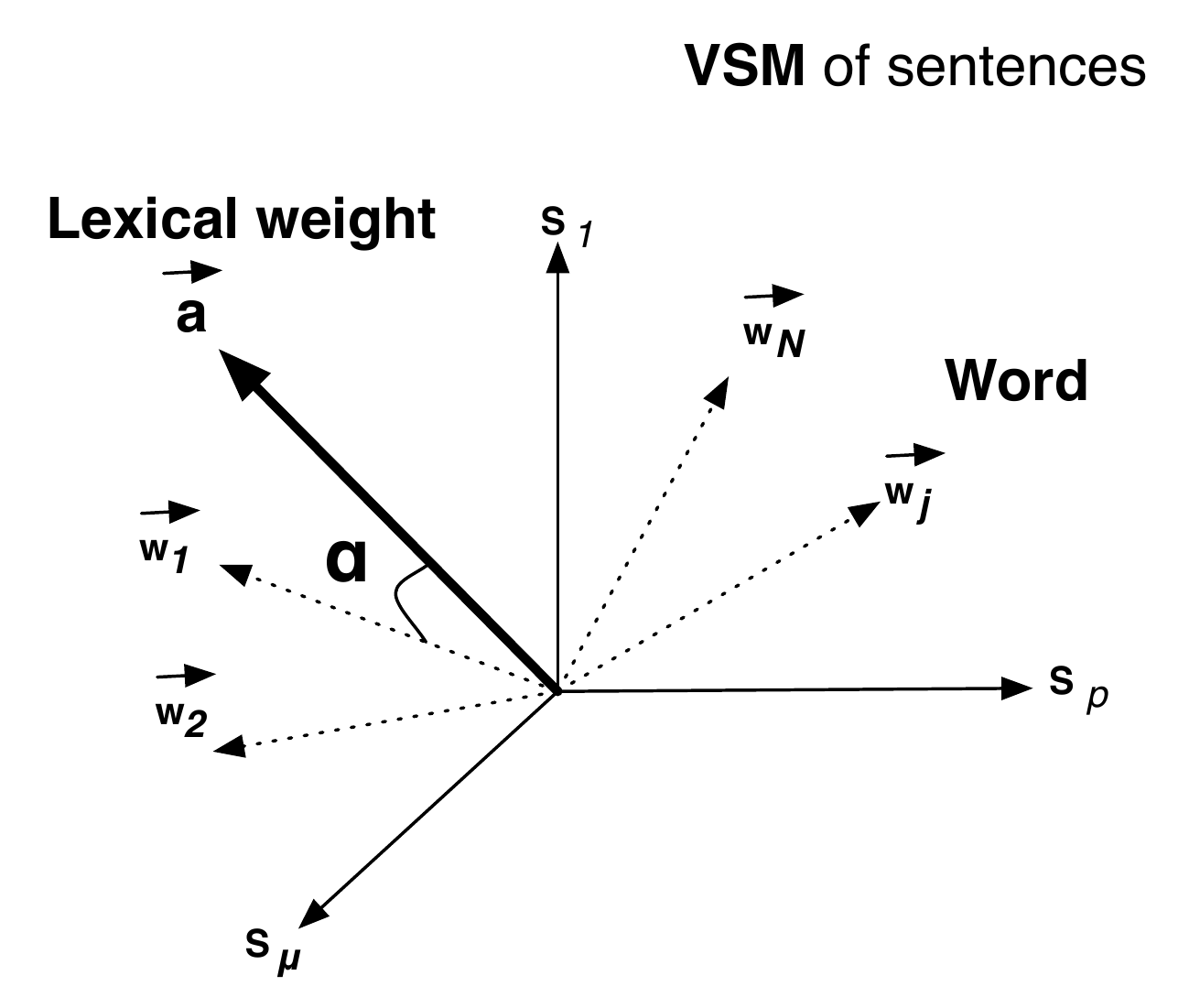}
 \caption{The ``lexical weight'' in a Vector Space Model of $p$ sentences.}
 \label{fig:artex_word}
\end{center}
\end{figure}

Next, a weight for each sentence is calculated using their proximity with the ``global topic'' and their ``lexical weight''. In Figure~\ref{fig:artex_word}, the ``lexical weight'' is represented in a VSM of $p$ sentences. Narrow angles indicate that words closest to the ``lexical weight'' should be important. Finally, the summary is generated concatenating the sentences with the highest scores following their order in the original document. Formally, ARTEX algorithm computes the score of each sentence by calculating the inner product between a sentence vector, an average pseudo-sentence vector (the ``global topic'') and an {\it average pseudo-word} vector(the``lexical weight''). Once the pre-processing is complete, a matrix $S_{[p×N]}$ ($N$ words and $p$ sentences) is created. Let $\overrightarrow{s_\mu}=(s_{\mu,1},s_{\mu,2},...,s_{\mu,N})$ be a vector of the sentence $\mu=1,2,...,p$. The {\it average pseudo-word} vector $\overrightarrow{a}=[a\mu]$ was defined as the average number of occurrences of $N$ words used in the sentence $\overrightarrow{s_\mu}$:
\begin{equation}
a_\mu=\frac{1}{N}\sum\limits_j s_{\mu,j}
\end{equation}

and the {\it average pseudo-sentence} vector $\overrightarrow{b}=[b_j]$ as the average number of occurrences of each word $j$ used through the $p$ sentences: 
\begin{equation}
b_j=\frac{1}{p}\sum\limits_\mu s_{\mu,j}
\end{equation}

The weight of a sentence $\overrightarrow{s_\mu}$ is calculated as follows:
\begin{align}
w(\overrightarrow{s_\mu})&=(\overrightarrow{s}\times \overrightarrow{b})\times \overrightarrow{a} \nonumber \\
					&=\frac{1}{N_p}\left( \sum\limits_{j=1}^N s_{\mu,j}\times b_j\right) \times a_\mu\ ; \mu=1,2,\dots,p
					\label{eq:sentenceWeight}
\end{align}

The $w(\bullet)$ computed by Equation~\ref{eq:sentenceWeight} must be normalized between the interval $[0,1]$. The calculation of $(\overrightarrow{s}\times\overrightarrow{b})$ indicates the proximity between the sentence $\overrightarrow{s_\mu}$ and the {\it average pseudo-sentence} $\overrightarrow{b}$. The product $(\overrightarrow{s}\times \overrightarrow{b})\times \overrightarrow{a}$ weight this proximity using the {\it average pseudo-word} $\overrightarrow{a}$. If a sentence $\overrightarrow{s_\mu}$ is near $\overrightarrow{b}$ and their corresponding element $a_\mu$ has a high value, therefore $\overrightarrow{s_\mu}$ will have a high score. Moreover, a sentence $\overrightarrow{s_\mu}$ far from a main topic (i.e. $\overrightarrow{s_\mu}\times \overrightarrow{b}$ is near $0$) or their corresponding element $a_mu$ has a low value, (i.e. $a_mu$ are near $0$), therefore $\overrightarrow{s_\mu}$ will have a low score.

It is not really necessary to divide the scalar product by the constant $\frac{1}{N_p}$, because the angle $\alpha$ between $\overrightarrow{b}$ and $\overrightarrow{s_\mu}$ is the same if $\overrightarrow{b}=\overrightarrow{b'}=\sum_\mu s_{\mu,j}$. The element $a_\mu$ is only a scale factor that does not modify $\alpha$~\cite{artex}:

\begin{align}
w(\overrightarrow{s_\mu})*&=\frac{1}{\sqrt{N^5p^3}}\left( \sum\limits_{j=1}^Ns_{\mu,j}\times b_j\right) \times a_\mu\ ;\mu=1,2,\dots,p
\label{eq:wStar}
\end{align}

The term $1/\sqrt{N^5p^3}$ is a constant value, and then $w(\bullet)$ (Equation~\ref{eq:sentenceWeight}) and $w(\bullet)*$ (Equation~\ref{eq:wStar}) are both equivalent.


This summarization system outperforms the CORTEX~\cite{Torres-Moreno2001} one with the FRESA~\cite{torres2010summary} measure. ARTEX is evaluated with several corpus such as the {\it Medecina Clinica}~\cite{artex}. ARTEX performance is then better than CORTEX on English, Spanish or French, which are the targeted languages in this study.

\section{Evaluation of LDA model quality}
\label{sec:SQM}

The previous section described different summarization systems to reduce the size of train corpus and to retain only relevant information contained into the train documents. This section proposes a set of metrics to evaluate the quality of topic spaces generated from summaries of the train documents. The first one is the perplexity. This score is the most popular one. We also propose to study another measure to evaluate the dispersion of each word into a given topic space. This measure is called the Jensen-Shannon ($\mathcal{JS}$) divergence.

\subsection{Perplexity}
\label{subsec:Perplexity}
Perplexity is a standard measure to evaluate topic spaces, and more generally a probabilistic model. A topic model $\mathcal{Z}$ is effective if it can correctly predict an unseen document from the test collection. The perplexity used in language modeling is monotonically decreasing in the likelihood of the test data, and is algebraically equivalent to the inverse of the geometric mean per-word likelihood. A lower perplexity score indicates better generalization performance~\cite{blei2003latent}:

\begin{equation}
perplexity(\mathcal{B})=exp\left\{-\frac{1}{N_{\mathcal{B}}}\sum\limits_{d=1}^{M} \log P({\bf w})\right\}
\end{equation}
with
\begin{equation}
N_{\mathcal{B}}=\sum\limits_{d=1}^{M} N_d
\end{equation}

where $N_{\mathcal{B}}$ is the combined length of all $M$ testing terms and $N_d$ is the number of words in the document $d$; $P({\bf w})$ is the likelihood that the generative model will be assigned to an unseen word ${\bf w}$ of a document $d$ in the test collection. The quantity inside the exponent is called the {\it entropy} of the test collection. The logarithm enables to interpret the entropy in terms of {\it bits} of information.

\subsection{Jensen-Shannon ($\mathcal{JS}$) divergence}
\label{subsec:KullbachLiebler}
The perplexity evaluates the performance of a topic space. Another important information is the distribution of words in each topic. The Kullback-Leibler divergence ($\mathcal{KL}$) estimates how much a topic is different from the $N$ topics contained in the topic model. This distribution is defined hereafter:
\begin{equation}
\label{eq:kl}
\text{$\mathcal{KL}(z_i,z_j)$}= \displaystyle\sum_{\substack{w\in \mathcal{A}}} p_i \log\frac{p_i}{p_j}
\end{equation}
where $p_i=P(w|z_i)$ and $p_j=P(w|z_j)$ are the probabilities that the word $w$ is generated by the topic $z_i$ or $z_j$. Thus, the symmetric $\mathcal{KL}$ divergence is named Jensen-Shannon ($\mathcal{JS}$) divergence metric. It is the mid-point measure between $\mathcal{KL}(z_i,z_j)$ and $\mathcal{KL}(z_j,z_i)$. $\mathcal{JS}$ is then defined with equation~\ref{eq:kl} as the mean of the divergences between $(z_i,z_j)$ and $(z_j,z_i)$ as:

\begin{eqnarray}
\label{eq:klSymetric}
\text{$\mathcal{JS}$}(z_i,z_j)	&=&\frac{1}{2}\left( \text{$\mathcal{KL}(z_i,z_j)$}+\text{$\mathcal{KL}(z_j,z_i)$} \right) \nonumber \\
		&=&\frac{1}{2} \sum_{w \in {\mathcal{A}}} \left(p_i \log\frac{p_i}{p_j} + p_j \log\frac{p_j}{p_i} \right)\ .
\end{eqnarray}

The $\mathcal{JS}$ divergence for the entire topic space is then defined as the divergence between each pair of topics composing the topic model $\mathcal{Z}$, defined in equation~\ref{eq:klSymetric} as:

\begin{eqnarray}
\label{eq:klFinal}
\text{$\mathcal{JS}$}(\text{$\mathcal{Z}$})  &=&\sum\limits_{z_i \in \text{$\mathcal{Z}$}} \sum\limits_{z_j \in \text{$\mathcal{Z}$}} \text{$\mathcal{JS}$}(z_i,z_j)  \nonumber \\ 
	    &=&\frac{1}{2}\sum\limits_{z_i \in \text{$\mathcal{Z}$}} \sum\limits_{z_j \in \text{$\mathcal{Z}$}} \sum_{w \in {\mathcal{A}}} p_i \log\frac{p_i}{p_j} + p_j \log\frac{p_j}{p_i}.
\end{eqnarray}

if $i=j \Rightarrow \log\frac{p_j}{p_i}=0$ ($log 1=0$). After defining the metrics to evaluate the quality of the model, the next section describes the experiment data sets and the experimental protocol.

\section{Experiments}
\label{sec:Experiments}

These summarization systems are used to compress and retain only relevant information into train text collection in each language. This section presents the experiments processed to evaluate the relevance and the effectiveness of the proposed system of fast and robust topic space building. First of all, the experimental protocol is presented, and then a qualitative analysis of obtained results is performed using evaluation metrics described in Section~\ref{sec:SQM}.

\subsection{Experimental protocol}
\label{subsec:data}

 In order to train topic spaces,  a large corpus of documents is required. Three corpus was used. Each corpus $\mathcal{C}$ is in a particular language (English, Spanish and French), and is composed of a training set $\mathcal{A}$ and a testing set $\mathcal{B}$. 
The corpus are composed of articles from Wikipedia. Thus, for each of the three languages, a set of 100,000 documents is collected. 90\% of the corpus is summarized and used to build topic spaces, while 10\% is used to evaluate each model (no need to be summarized).

Table~\ref{tbl:dataBaseAll} shows that the latin languages (French and Spanish) have a similar size (a difference of less than 4\% is observed), while the English one is bigger than the others (English text corpus is $1.37$ times bigger than French or Spanish corpus). In spite of the size difference of corpus, both of them have more or less the same number of words and sentences in an article. We can also note that the English vocabulary  size is roughly the same ($15\%)$ than the latin languages. Same observations can be made in Table~\ref{tbl:dataBasePerDocuments}, that presents statistics at document level (mean on the whole corpus). In next section, the outcome of this fact is seen during the perplexity evaluation of topic spaces built from English train text collection. 

\begin{table}[h!]
\caption{\label{tbl:dataBaseAll}Dataset statistics of the Wikipedia corpus.}
\centering
\scalebox{0.9}{
\begin{tabular}{|c | c c c|}
\hline
{\bf Language} & {\bf \#Words} & {\bf \#Unique Words} & {\bf \#Sentences}\\
\hline
English & 30,506,196& 2,133,055& 7,271,971\\
\hline
Spanish & 23,742,681& 1,808,828& 5,245,507\\
\hline
French  &25,545,425& 1,724,189& 5,364,825\\
\hline
\end{tabular}
}
\end{table}

\begin{table}[h!]
\caption{\label{tbl:dataBasePerDocuments}Dataset statistics per document of the Wikipedia corpus.}
\centering
\scalebox{0.9}{
\begin{tabular}{|c | c c c|}
\hline
{\bf Language} & {\bf \#Words} & {\bf \#Unique Words} & {\bf \#Sentences}\\
\hline
English & 339& 24& 81\\
\hline
Spanish & 264& 20& 58\\
\hline
French  &284& 19& 56\\
\hline
\end{tabular}
}
\end{table}


As set of topic spaces is trained to evaluate the perplexity and the Jensen-Shannon ($\mathcal{JS}$) scores for each language, as well as the processing time to summarize and compress documents from the train corpus. Following a classical study of LDA topic spaces quality~\cite{rosen2004author}, the number of topics by model is fixed to $\{5,10,50,100,200,400\}$. These topic spaces are built with the MALLET toolkit~\cite{McCallumMALLET}.

\subsection{Results}
\label{subsec:Results}

\begin{figure*}[!htpb]
\begin{center} 
\hspace{15mm}
\includegraphics[scale=0.4]{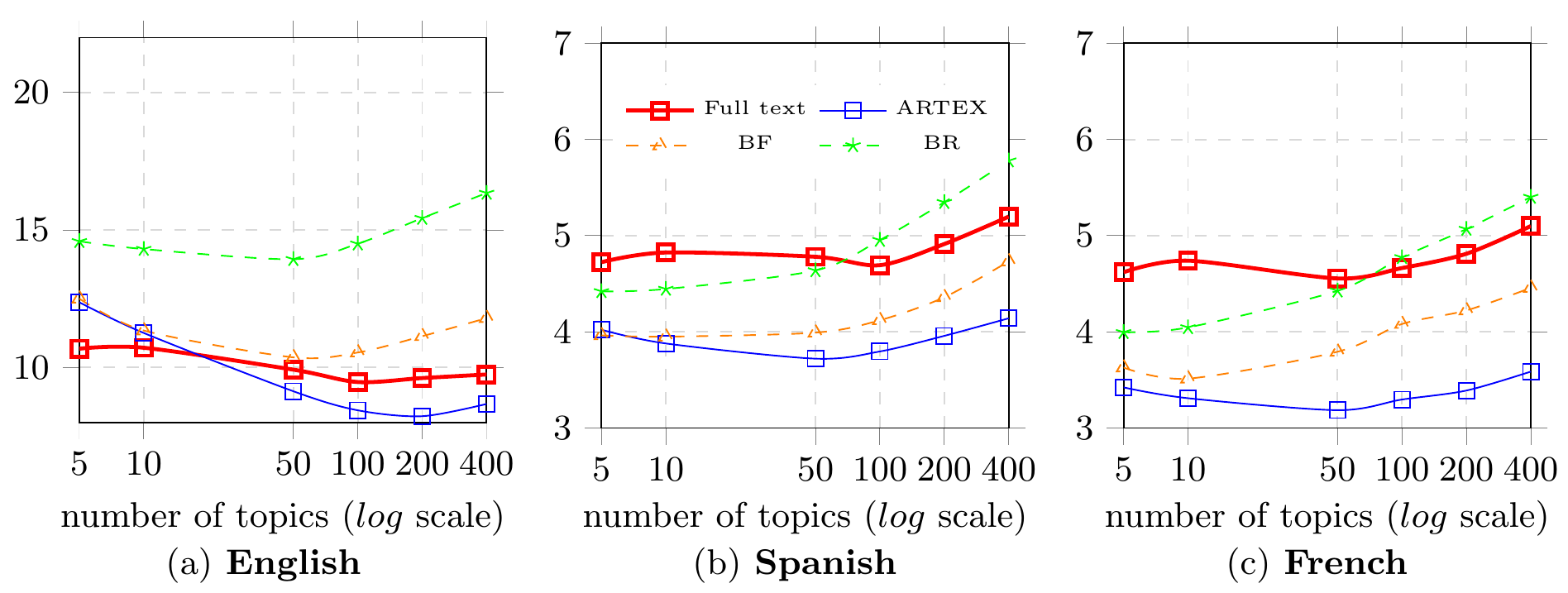} 
\end{center} 
\caption{Perplexity ($\times 10^{-3}$) by varying the number of topics for each corpus.} 
\label{exp:GraphiquesPerplexity_topics}
\end{figure*}

\begin{figure*}[!htpb]
\begin{center} 
\includegraphics[scale=0.4]{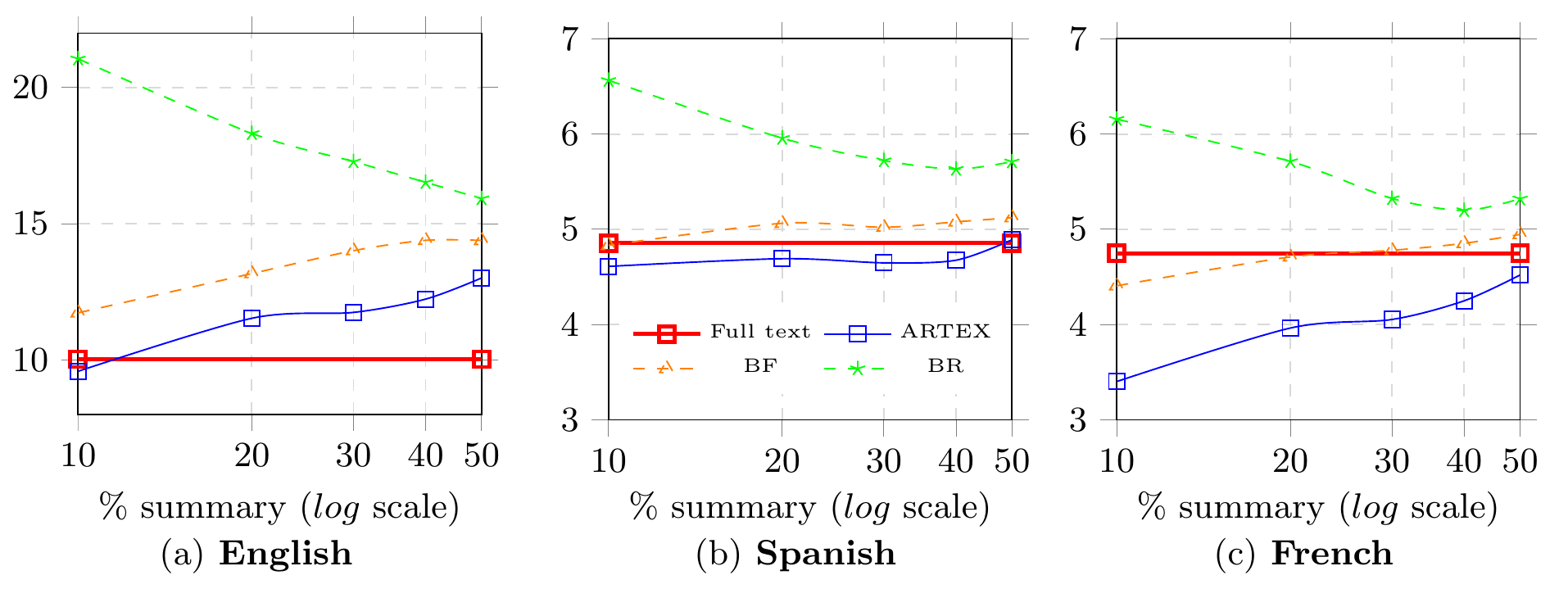} 
\end{center} 
\caption{Perplexity ($\times 10^{-3}$) by varying the \% summary for each corpus.} 
\label{exp:GraphiquesPerplexity_sum}
\end{figure*}

The experiments conducted in this paper are topic-based concern. Thus, each metric proposed in Section~\ref{sec:SQM} (Perplexity and $\mathcal{JS}$) are applied for each language (English, Spanish and French), for each topic space size ($\{5,10,50,100,200,400\}$), and finally, for each compression rate during the summarization process (10\% to 50\% of the original size of the documents). Figures~\ref{exp:GraphiquesPerplexity_topics} and~\ref{exp:GraphiquesPerplexity_sum} present results obtained by varying the number of topics (Figure (a) to (c)) and the percentage of summary (Figure~\ref{exp:GraphiquesPerplexity_sum}), respectively for perplexity and Jensen-Shannon ($\mathcal{JS}$) measures. Results are computed with a mean among the various topic spaces size and a mean among the different reduced summaries size. 
Moreover, each language was study separately to point out differences of topic spaces quality depending on the language.


\newcommand{\width}{5.2cm}
\newcommand{\height}{5cm}

\section{Discussions}
\label{sec:Discussion}

The results reported in Figures~\ref{exp:GraphiquesPerplexity_topics} and~\ref{exp:GraphiquesPerplexity_sum} allow us to point out a first general remark, already observed in section~\ref{subsec:data}: the two latin languages have more or less the same tendencies. This should be explained by the root of these languages, which are both latins. 

\begin{figure*}[!htb]
\begin{center} 
\hspace{-15mm}
\includegraphics[scale=0.4]{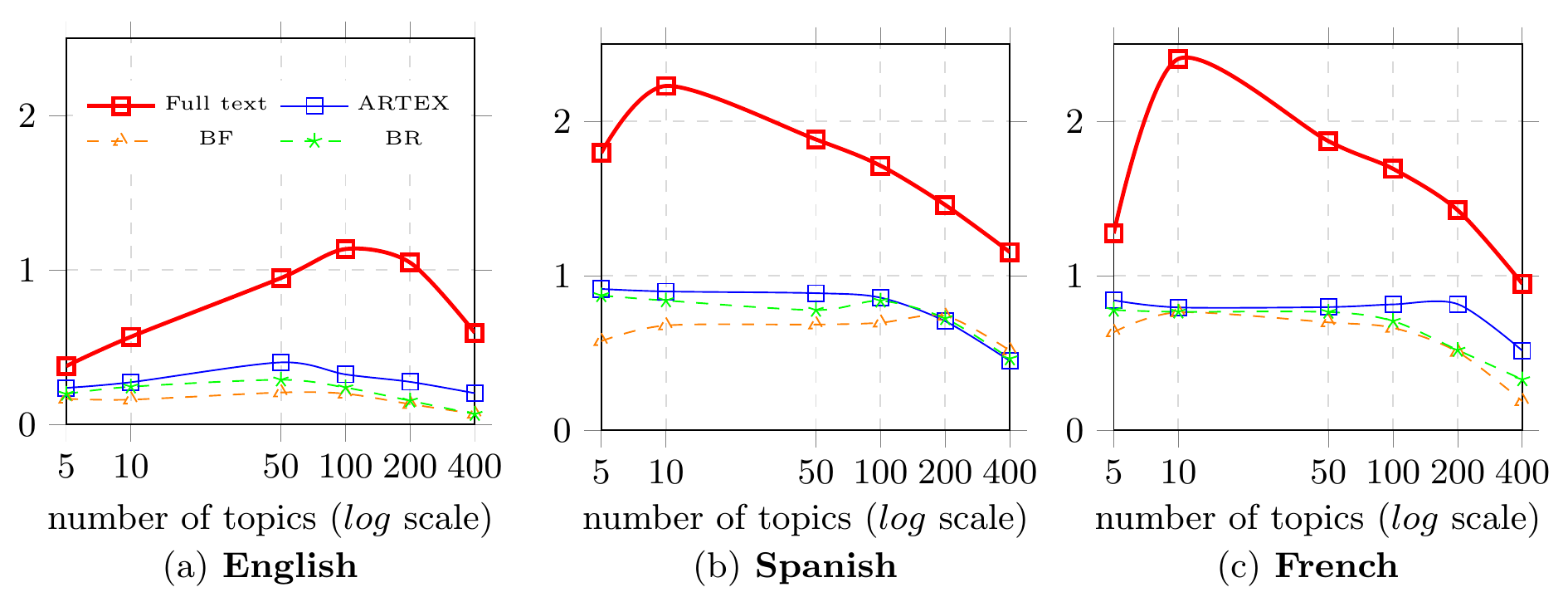} 
\end{center} 
\caption{Jensen-Shannon ($\times 10^{3}$) measure by varying the number of topics for each corpus.} 
\label{exp:GraphiquesKL_topics}
\end{figure*}

\begin{figure*}[!htb]
\begin{center} 
\hspace{-15mm}
\includegraphics[scale=0.4]{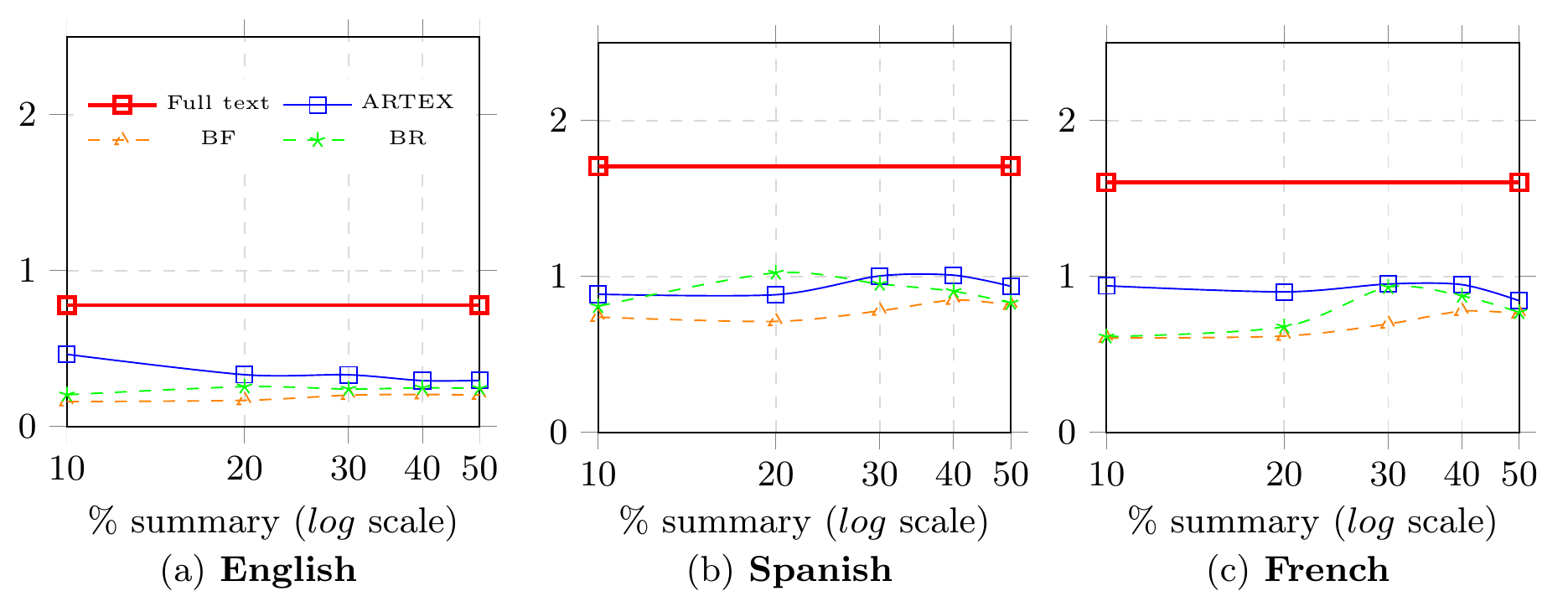} 
\end{center} 
\caption{Jensen-Shannon ($\times 10^{3}$) measure by varying the \% summary for each corpus.} 
\label{exp:GraphiquesKL_sum}
\end{figure*}

Figure~\ref{exp:GraphiquesPerplexity-topics} shows that the Spanish and French corpus obtain a perplexity between 3,000 and 6,100 when the number of classes in the topic space varies. Another observation is that, for these two languages, topic spaces obtained with summarized documents, outperform the ones obtained with complete documents when at least 50 topics are considered (Figures~\ref{exp:GraphiquesPerplexity_topics}-b and -c). The best system for all languages is ordered in the same way. Systems are ordered from the best to the worst in this manner: ARTEX, BF (this fact is explained in the next part and is noted into $\mathcal{JS}$ measure curves in Figures~\ref{exp:GraphiquesKL_topics} and~\ref{exp:GraphiquesKL_sum}), and then BR. If we considerer a number of topics up to $50$, we can note that the topic spaces, from full text documents ({\it i.e.} not summarized) with an English text corpus, obtain a better perplexity (smaller) than documents processed with a summarization system, that is particularly visible into Figures~\ref{exp:GraphiquesPerplexity_sum}.

To address the shortcoming due to the size of the English corpus (1.37 times bigger than latin languages), the number of topics contained into the thematic space has to be increased to effectively disconnect words into topics. In spite of moving up, the number of topics move down the perplexity of topic spaces for all summarization systems (except random baseline (RB)), the perplexity obtained with the English corpus being higher than those obtained from the Spanish and French corpus.

Among all summarization systems used to reduce the documents from the train corpus, the baseline first (BF) obtains good results for all languages. This performance is due to the fact that BF selects the first paragraph of the document as a summary: when a Wikipedia content provider writes a new article, he exposes the main idea of the article in the first sentences. Furthermore, the rest of the document relates different aspects of the article subject, such as historical or economical details, which are not useful to compose a relevant summary. Thus, this baseline is quite hard to outperform when the documents to summarize are from encyclopedia such as Wikipedia.


The random baseline (RB) composes its summary by randomly selecting a set of sentences in an article. This kind of system is particularly relevant when the main ideas are disseminated in the document such as a blog or a website. This is the main reason why this baseline did not obtain good results except for $\mathcal{JS}$ divergence measure (see Figures~\ref{exp:GraphiquesKL_topics} and~\ref{exp:GraphiquesKL_sum}). This can be explained by the fact that this system selects sentences at different places, and then, selects a variable set of words. Thus, topic spaces from these documents contain a variable vocabulary. The $\mathcal{JS}$ divergence evaluates how much a word contained in a topic is discriminative, and allows to distinguish this topic from the others that compose the thematic representation.

Figures~\ref{exp:GraphiquesKL_topics} and~\ref{exp:GraphiquesKL_sum} also show that Jensen-Shannon ($\mathcal{JS}$) divergence scores between topics obtained a similar performance order of summarization systems for all languages corpus. Moreover, full text documents always outperform all topic spaces representation for all languages and all summary rates. The reason is that full text documents contain a larger vocabulary, and $\mathcal{JS}$ divergence is sensitive to the vocabulary size, especially when the number of topics is equal for summarized and full text documents. This observation is pointed out by Figures~\ref{exp:GraphiquesKL_sum}-b and~-c where the means among topic spaces for each summary rate of full text documents are beyond all summarization systems. Last points of the curves show that topic spaces, with a high number of topics and estimated from summaries, do not outperform those estimated from full text documents, but become more and more closer to these ones: this confirms the original idea that have motivated this work.

Tables~\ref{exp:GraphiquesTime1} and~\ref{exp:GraphiquesTime2} finally present the processing time, in seconds, by varying the number of topics for each language corpus, respectively with the full text and the summarized documents.
We can see that processing time is saved when topic spaces are learned from summarized documents. Moreover, tables show that the processing times follow an exponential curve, especially for the full text context. For this reason, we can easily imagine the processing time that can be saved using summaries instead of the complete documents, which inevitably contain non informative and irrelevant terms.

A general remark is that the best summarization system is ARTEX, but if we take into account the processing time during the topic space learning, the baseline first (BF) is the best agreement. Indeed, if one want to find a common ground between a low perplexity, a high $\mathcal{JS}$ divergence between topics and a fast learning process, the BF method should be chosen. 

\begin{table}[h!]
\caption{Processing time (in seconds) by varying the number of topics for each corpus.} 
\label{exp:GraphiquesTime1}
\centering
\scalebox{0.9}{
\begin{tabular}{|c | c c c|}
\hline
{\bf System} & \multicolumn{3}{c|}{\bf Language}\\
\hline
{\bf Full Text} & {\bf English} & {\bf Spanish} & {\bf French}\\
\hline
5 &  1,861& 1,388 & 1,208\\
10 &  2,127& 1,731 & 1,362\\
50 &  4,194& 2,680 & 2,319\\
100 &  5,288& 3,413 & 3,323\\
200 & 6,364 & 4,524 & 4,667\\
400 & 8,654 & 6,625 & 6,751\\
\hline
\end{tabular}
}
\end{table}

\begin{table*}[htpb!]
\caption{Processing time (in seconds) by varying the number of topics for each corpus.} 
\label{exp:GraphiquesTime2}
\centering
\scalebox{0.75}{
\begin{tabular}{|c | c c c || c | c c c || c | c c c|}
\hline
{\bf System} & \multicolumn{3}{c||}{\bf Language}& {\bf System} & \multicolumn{3}{c||}{\bf Language} & {\bf System} & \multicolumn{3}{c|}{\bf Language}\\
\hline
{\bf ARTEX} & {\bf English} & {\bf Spanish} & {\bf French} & {\bf BR} & {\bf English} & {\bf Spanish} & {\bf French} & {\bf BF} & {\bf English} & {\bf Spanish} & {\bf French}\\
\hline
5 & 514 & 448 & 394 & 5 & 318 & 265 & 238 & 5 &  466 & 301& 276\\
10 & 607 & 521 & 438 & 10 & 349 & 298 & 288 & 10 &  529 & 348 & 317\\
50 & 1,051 & 804 & 709 & 50 & 466 & 418 & 465 & 50 &  1031 & 727 & 459\\
100 & 1,565 & 1,303 & 1,039 & 100 & 652 & 602 & 548 & 100 &  1,614& 737 & 680\\
200 & 2,536 & 2,076 & 1,573 & 200 & 919 & 863 & 838 & 200 & 2,115 & 814 & 985\\
400 & 3,404 & 2,853& 2,073 & 400 & 1,081 & 988 & 978  & 400 & 2,784 & 1,448 & 988\\
\hline
\end{tabular}
}
\end{table*}

\section{Conclusions}
\label{sec:conclusionsAndFutureWork}

In this paper, a qualitative study of the impact of documents summarization in topic space learning is proposed. The basic idea that learning topic spaces from compressed documents is less time consuming than learning topic spaces from the full documents is noted. The main advantage to use the full text document in text corpus to build topic space is to move up the semantic variability into each topic, and then increase the divergence between these ones. Experiments show that topic spaces with enough topics size have more or less (roughly) the same divergence. 

Thus, topic spaces with a large number of topics, {\it i.e.} suitable knowing the size of the corpus (more than 200 topics in our case), have a lower perplexity, a better divergence between topics and are less time consuming during the LDA learning process. The only drawback of topic spaces learned from text corpus of summarized documents disappear when the number of topics comes up suitable for the size of the corpus whatever the language considered.

\bibliographystyle{splncs}
\bibliography{mohamed_morchid_biblio}

\end{document}